\documentclass[conference]{IEEEtran}
\IEEEoverridecommandlockouts
\usepackage{cite}
\usepackage{amsmath,amssymb,amsfonts}
\usepackage{algorithmic}
\usepackage{graphicx}
\usepackage{textcomp}
\usepackage{xcolor}
\usepackage[pangram]{blindtext}

\def\BibTeX{{\rm B\kern-.05em{\sc i\kern-.025em b}\kern-.08em
    T\kern-.1667em\lower.7ex\hbox{E}\kern-.125emX}}

\usepackage[utf8x]{inputenc}

\makeatletter
\def\endthebibliography{%
  \def\@noitemerr{\@latex@warning{Empty `thebibliography' environment}}%
  \endlist
}

 \let\old@ps@headings\ps@headings
 \let\old@ps@IEEEtitlepagestyle\ps@IEEEtitlepagestyle
 \def\confheader#1{%
 \def\ps@headings{%
 \old@ps@headings%
 \def\@oddhead{\strut\hfill#1\hfill\strut}%
 \def\@evenhead{\strut\hfill#1\hfill\strut}%
 }%
 \def\ps@IEEEtitlepagestyle{%
 \old@ps@IEEEtitlepagestyle%
 \def\@oddhead{\strut\hfill#1\hfill\strut}%
 \def\@evenhead{\strut\hfill#1\hfill\strut}%
 }%
 \ps@headings%
 }
 \makeatother
 
\confheader{Preprint submitted to the 35\textsuperscript{th} International Conference on Software Maintenance and Evolution (ICSME'19)}
 
\begin{document}

\title{Tracy: A Business-driven Technical Debt Prioritization Framework}

\author{
    \IEEEauthorblockN{Rodrigo Rebouças de Almeida\IEEEauthorrefmark{1}\IEEEauthorrefmark{3}, Christoph Treude\IEEEauthorrefmark{2}, Uir\'{a} Kulesza\IEEEauthorrefmark{3}}
\IEEEauthorblockA{\IEEEauthorrefmark{1}Federal University of Paraíba - UFPB, Rio Tinto, Paraíba, Brazil}
\IEEEauthorblockA{\IEEEauthorrefmark{2}University of Adelaide, Adelaide, Australia}
\IEEEauthorblockA{\IEEEauthorrefmark{3}Federal University of Rio Grande do Norte - UFRN, Natal, Brazil}
rodrigor@dcx.ufpb.br, christoph.treude@adelaide.edu.au,  uira@dimap.ufrn.br}

\maketitle

\begin{abstract}
Technical debt is a pervasive problem in software development. Software development teams have to prioritize debt items and determine whether they should address debt or develop new features at any point in time. This paper presents ``Tracy'', a framework for the prioritization of technical debt using a business-driven approach built on top of business processes. The current stage of the proposed framework is at the beginning of the third phase of Design Science Research, which is usually divided into the phases of exploration, engineering, and evaluation. The exploration and engineering phases involved the participation of 49 professionals from 12 different groups of three companies. The initial evaluation shows that the presented framework is coherent in its structure and that its results contribute to business-driven decision making on technical debt prioritization.
\end{abstract}

\section{Introduction}

Technical debt is a pervasive problem in software development and evolution that is introduced when teams take a shortcut to gain short-term benefits at the cost of making future changes more expensive or impossible~\cite{Kruchtenbook2019}. Management and business factors are the leading causes of technical debt~\cite{rios2019}, and many researchers have pointed out that research should focus more on the business aspects of technical debt~\cite{AMPATZOGLOU:2015,terese2019}. 

In this paper, we present a business-driven technical debt prioritization framework, called ``Tracy'', that prioritizes technical debt considering how IT assets (IT systems which create business value) support a company's business processes. Tracy uses business metrics to support the decision making and has two major benefits: (1) it encourages different stakeholders to consider and identify the business metrics that support decision making about technical debt, and (2) it provides a prioritization mechanism that has the potential to be applied in different business and development contexts.

The current stage of the proposed framework is at the beginning of the third phase of Design Science Research (DSR)~\cite{dsr.book}, which is usually divided into the three phases of exploration, engineering, and evaluation. The exploration and engineering phases involved the participation of 49 professionals from 12 different groups of three companies.

To the best of our knowledge, this is the first paper which proposes a technical debt prioritization framework considering business processes and business metrics. The initial evaluation shows that the presented framework is coherent in its structure, and that its results contribute to business-driven decision making on technical debt prioritization.
\section{Methodology}

Due to the importance of considering business aspects when managing and prioritizing technical debt, we are conducting Design Science Research (DSR) to develop a solution for the following design goal/problem statement~\cite{dsr.book}:
\textbf{Improve technical debt prioritization by designing a business-oriented decision-making framework to promote the alignment between technical decisions and business expectations.}

The designed solution relies on the analysis of information collected over six months and 22 meetings (interviews and focus groups) with seven different groups in two companies, with engineering involving an additional company. Additionally, the first author participated as an observer in 12 events (sprint plannings, sprint reviews, incidents, and decision making), where he was able to witness technical debt creation, identification, payment, and business impact.

To conceive a solution for the design problem which would apply to more than one company, the DSR stakeholders were composed of a set of 14 groups of participants from three companies. The groups included 43 professionals: 10 with pure-business roles, 9 with management roles, 6 with technical leadership roles, and 18 with technical roles. No group was aware of the research activities conducted with other groups, and all management and business professionals had more than ten years of professional experience. 

The companies have a typical IT organization, with development teams, operations, and use of cloud infrastructure to deliver their services. Two of them provide solutions to the Fintech industry, and the other is a global software consulting company. None of the companies employed a systematic technical debt management approach, often storing technical debt items as ``improvements'' in the backlog instead.

\begin{table}
\begin{center}
\caption{Groups (A1, A2, ...) and research activities in each DSR phase. In each pair \textit{x-y}, \textit{x} is the maximum number of participants, and \textit{y} is the number of iterations with the group.}
\includegraphics[width=.9\columnwidth]{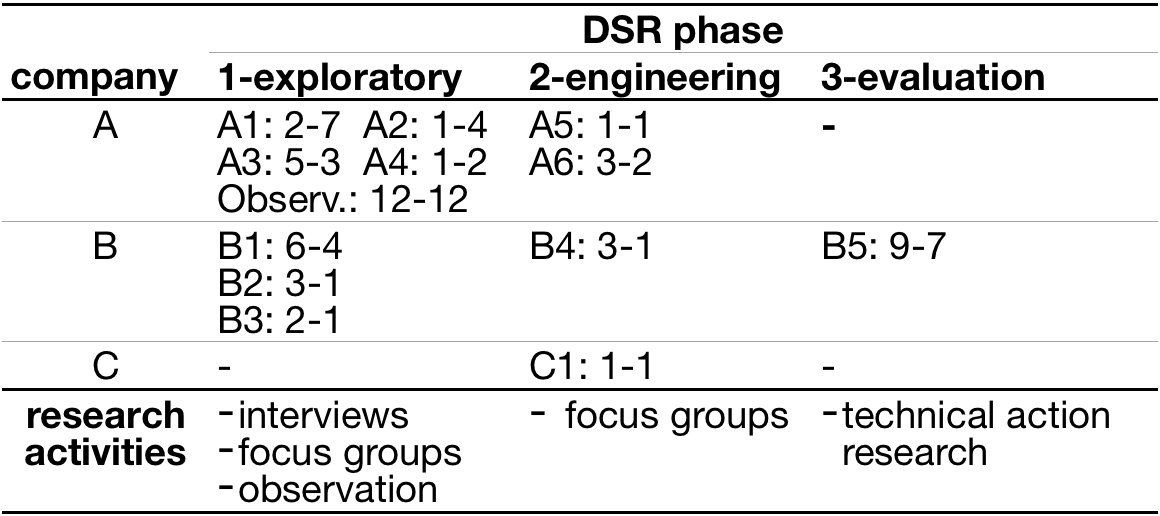}
\vspace{-.5cm}
\label{table.dsr.phases}
\end{center}
\end{table}

In each DSR phase, we conducted a set of research activities with selected groups, focusing on different objectives. The phases define the main objectives and the research focus but do not limit the possibility of improving the results of one phase in another, e.g., while designing the solution in phase~2, our understanding of the problem was refined as a result of ongoing discussions related to the solution design.

Table~\ref{table.dsr.phases} presents the number of groups and the research activities conducted in each phase. All research activity was captured in research log books and audio recordings.

\paragraph{Exploration Phase}

In the exploration phase, the objective was to understand the constructs (concepts, relations, rules, and motivations) related to technical debt decision making, considering business and management perspectives. The groups that we worked with in this phase had a majority of business and management participants. We conducted interviews, focus groups, and observations~\cite{book.sage} with eight groups, until we reached saturation of the constructs. We also evaluated the business-driven approach to prioritize technical debt in two case studies~\cite{icsme2018}. The approach was a first step in terms of the work extent and the level of business impact measurement, and a building block for the solution proposed in the next phase.

\paragraph{Engineering Phase}

The engineering phase started with a first-version prescriptive framework conceived on top of the information from the previous phase and elements from our previous work~\cite{icsme2018}. Then, we defined key requirements for the solution and iterated over five groups in three companies (see Table~\ref{table.dsr.phases}, phase 2). The groups in this phase had participants with business/management and technical background since the framework needed the input of both profiles. This phase was conducted using focus groups~\cite{book.sage} where the first author observed participants using a version of the solution. After each iteration, the framework was reviewed, improved, and presented to the same group and to another one. The end of this phase was triggered when the groups did not have anything to add to the framework.
\paragraph{Evaluation Phase}

The objective of the evaluation phase was to verify if the solution met the requirements specified in the previous phase, if it solved the design problem, and if the concepts, relationships, artifacts, and prioritization criteria were valid. The evaluation was conducted using Technical Action Research (TAR)~\cite{dsr.book} to enable stakeholders to learn about the effects of the framework in practice.

\section{A Business-Driven Framework for TDM}

\begin{figure}[ht]
\begin{center}
\includegraphics[width=1\columnwidth]{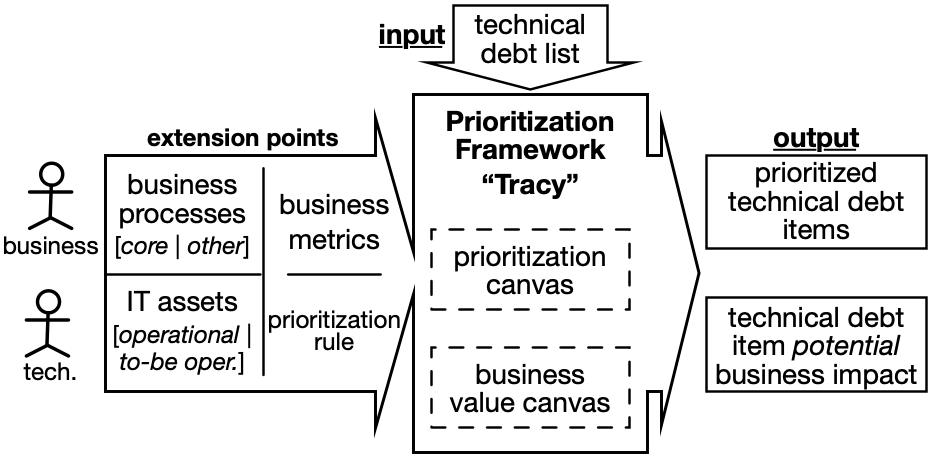}
\caption{Components of the technical debt prioritization framework}
\label{fig_tracy}
\vspace{-.6cm}
\end{center}
\end{figure}

Figure~\ref{fig_tracy} shows \textit{Tracy}, our business-driven technical debt prioritization framework.
It is based on the assumption that business processes are a way to identify business values~\cite{bpmbookdumas}, and receives as input a \textit{technical debt list} and produces as output a \textit{prioritized set of technical debt items} and their \textit{potential business impact}.

The prioritization is done using the following information, shown as framework \textit{extension points} in Figure~\ref{fig_tracy}:

\begin{itemize}
  \item \textbf{Business processes}: a list of business processes (BPs) supported by IT assets. The processes are classified as \textit{core/support} or \textit{other};
  \item \textbf{IT assets}: a list of IT assets that support the business processes;
  \item \textbf{Prioritization rule}: a rule to prioritize technical debt considering the impact of the IT assets on their supported business processes;
  \item \textbf{Business metrics} related to each business process and IT asset.
\end{itemize}

The output \textit{``prioritized technical debt items''} is a set of technical debt items grouped by the relationship between their affected IT assets and their supported business processes, based on the prioritization rule described in the next Section.

The solution relies on the following two elements and a procedure to set them up: (1) a prioritization canvas that drives a first prioritization step, based on the relationship between the IT assets and their supported business processes; and (2) a business-value canvas to support the measurement of technical debt items' potential business impact through a set of business metrics.

\subsection{Setup procedure}
\label{section.procedure}

To instantiate the framework, three steps have to be followed which we describe in this section.

\label{section.identify.assets}
\subsubsection{Define the IT assets which will be managed}

An IT asset is a software system, service, or application that creates business value (i.e., revenue or business opportunity) or may incur business loss (i.e., cost or penalty). IT assets are classified as \textit{operational} or \textit{to-be operational}.
IT asset is an abstract concept which may refer to a set of systems and subsystems or a particular service. For example, in the case of one of the groups we worked with for this research, an IT asset corresponded to a set of micro-service endpoints.

\label{section.identify.ci}
\subsubsection{Identify the configuration items (CIs) and their interdependencies and relate them to the IT assets}

\begin{figure}
\begin{center}
\includegraphics[width=.8\columnwidth]{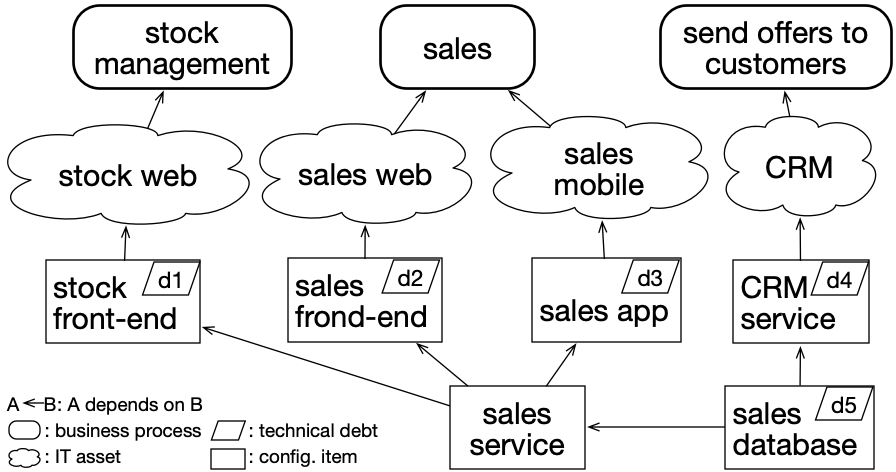}
\caption{Example of the relationship between configuration items (affected by technical debt d1-d5), their supported IT assets and the business processes.}
\vspace{-.5cm}
\label{fig_ci}
\end{center}
\end{figure}

The configuration items are the subsystems, modules, databases, and software infrastructure which are affected by technical debt and directly or indirectly support IT assets. Their identification must be conducted in consultation with IT stakeholders. The appropriate level of abstraction will depend on the type of technical debt, e.g., architectural debt will better relate to systems and modules, while unit-test debt would relate better to classes and packages. Figure~\ref{fig_ci} shows an example of how technical debt items (d1-d5) are related to configuration items, along with their dependencies and support for IT assets and business processes. In the example, the technical debt \textit{d3} affects the \textit{Sales App} configuration item, which supports the \textit{SalesMobile} IT asset, which, finally, supports the \textit{Sales} business process.

\label{section.identify.bp}
\subsubsection{Identify the business processes supported by the IT assets} 

\textit{``core''} business processes are those which deliver the main value to the business, \textit{``support''} processes are processes which provide support to the core processes, e.g., the payment business process which supports sales, and  \textit{``other''} processes are management or peripheral processes, e.g., \textit{``stock management''} or \textit{``send offers to customers''}~\cite{bpmbookdumas}.

\label{section.canvas}
\subsection{The Prioritization canvas}

\begin{figure}
\begin{center}
\includegraphics[width=.8\columnwidth]{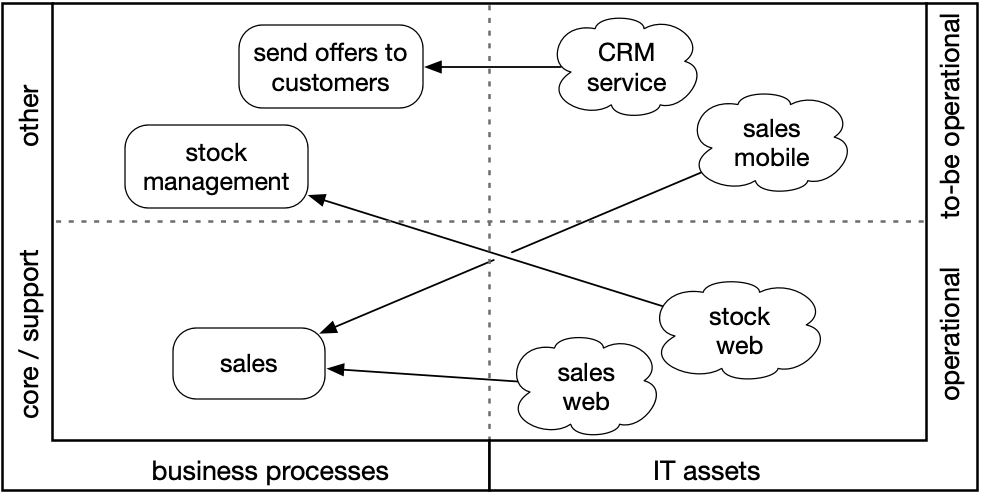}
\caption{Technical debt business-driven prioritization canvas}
\vspace{-.5cm}
\label{fig_prioritization_canvas}
\end{center}
\end{figure}

The \textit{Prioritization Canvas} (Figure~\ref{fig_prioritization_canvas}) is a board to visualize the relationship between IT assets and their supported business processes. It is composed of four quadrants, where the business processes and their supported IT assets are arranged according to their types and states. On the left side are the business processes, categorized into \textit{core/support} and \textit{others}. On the right side are the IT assets, grouped into \textit{operational} and \textit{to-be operational}. The arrows express dependencies between them, e.g., the \textit{Sales} business process depends on the \textit{Sales web} system.
\begin{table}
\begin{center}
\caption{Example of technical debt priority considering the IT assets and their supported business processes}
\includegraphics[width=.7\columnwidth]{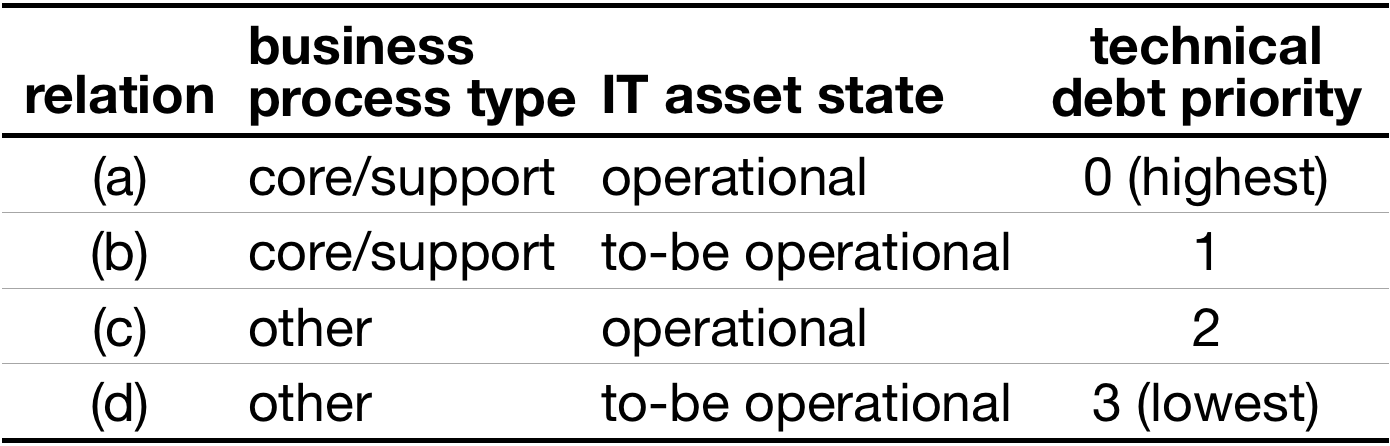}
\vspace{-.5cm}
\label{table_priority}
\end{center}
\end{table}

The prioritization of technical debt items follows the relationship between IT assets and their supported business processes. Each company or project must decide what will guide this first-step prioritization by defining a priority for each combination of business process type and IT asset operational state. Table~\ref{table_priority} shows an example of a prioritization rule where technical debt items that affect IT assets which support \textit{core/support} business processes have a higher priority than others.
It is important to highlight that each company or even each project within the same company may have different prioritization rules. For instance, we observed a case where a business unit had an operational legacy system that supported a core business process, and a new system (to-be operational) that was being developed to substitute the legacy one. In this case, the head of the unit decided that technical debt from the new system should have a higher priority than the one from the legacy system.

\label{section.valueCanvas}
\subsection{The Business-value canvas}

\begin{figure}
\begin{center}
\includegraphics[width=1\columnwidth]{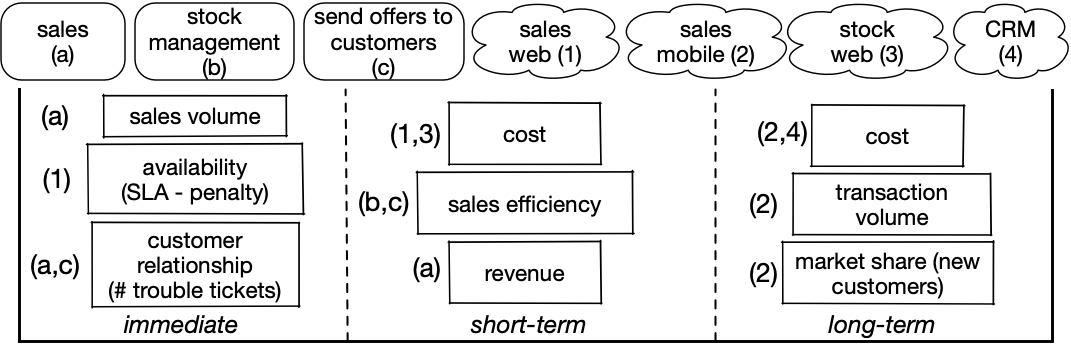}
\caption{Business-value canvas}
\vspace{-.5cm}
\label{fig_businessvalue_canvas}
\end{center}
\end{figure}

The business-value canvas, where each business process and IT asset is related to metrics which may have immediate, short-term, or long-term business impact, is shown in Figure~\ref{fig_businessvalue_canvas}. It is a tool to help stakeholders identify and classify the business value created by business processes and IT assets. Depending on the company or project strategy, the time periods can be different from `immediate, short-term, and long-term'---these periods were adequate for the scenarios considered in this work. The framework may instantiate more than one canvas, e.g., one for each business process and IT asset or one for each pair of asset and process.

The business-value canvas aims at determining what is the potential immediate, short-term, and long-term business impact of technical debt which affects an IT asset.

To identify the metrics, one must consider technical debt as a risk factor which affects the business value~\cite{allman2012}, i.e., for each business process and IT asset, one must identify how they affect the business, objectively. For example, in Figure~\ref{fig_businessvalue_canvas}, \textit{a technical debt which affects} the \textit{Sales} BP has an immediate potential business impact on \textit{Sales volume}, and on the \textit{Customer Relationship}. It can also have impact on the Revenue (in short term), and on the planned increased sales volume (in long term). The Sales BP is supported by the \textit{Sales web} system, where technical debt can have immediate impact on \textit{availability} and may have impact on the \textit{Cost} in short-term.

This canvas has three benefits: (i) the exercise that it promotes when stakeholders are completing it---during our research, participants thought about their business, their relationship with customers, and operational versus strategic impact; (ii) the establishment of concrete and standard metrics as a basis for decisions of various stakeholders, thus enabling IT leaders to debate with business stakeholders based on a common understanding of the business impact of technical debt; and (iii) the provision of real-time information about how technical debt can affect business. 

The exact metrics will depend on a company's business model and how it can be affected by, e.g., its customers, contracts, SLAs, market competitiveness, and strategic or operational plans. 

\subsection{Technical Debt Prioritization}

\begin{table*}[ht]
\begin{center}
\caption{Technical debt prioritization example}
\centerline{\includegraphics[width=0.85\textwidth]{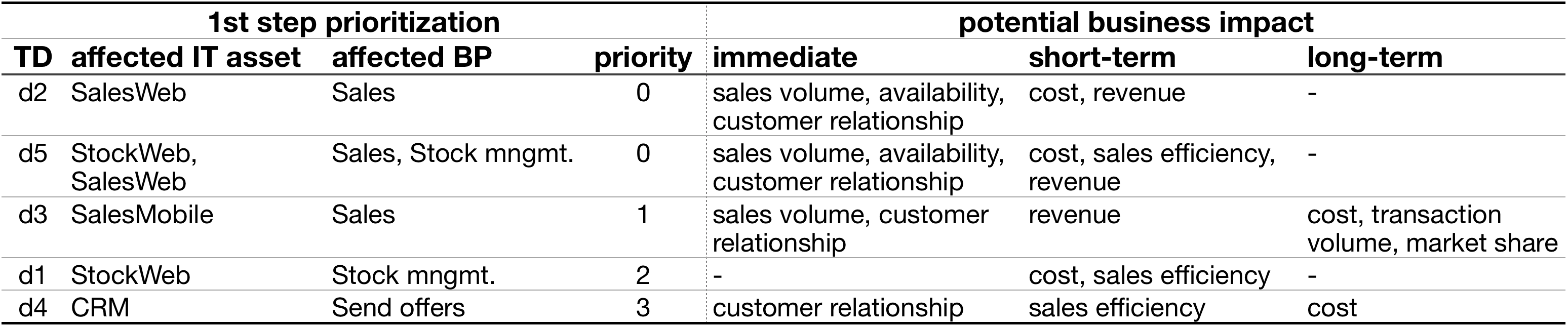}}
\vspace{-0.8cm}
\label{table_example_priority}
\end{center}
\end{table*}

We show an example of how the prioritization happens and how business metrics are organized to support the decision making about technical debt prioritization. Table~\ref{table_example_priority} shows the prioritization of the five technical debt items (d1-d5) shown in Figure~\ref{fig_ci}. The first-step prioritization considers the relationship between business processes and their supported IT assets using the prioritization canvas (Figure~\ref{fig_prioritization_canvas}). The potential business impact is retrieved from the business-value canvas (Figure~\ref{fig_businessvalue_canvas}). 

Table~\ref{table_example_priority} shows that technical debt items d2 and d5 have the same highest priority: zero. These items should be the focus of decision making about ``Which technical debt should we focus on now?'' Since d2 affects operational services that impact the sales business process, it should be investigated, e.g., by verifying what type of technical debt it is and if it has immediate, short-term, or long-term potential business impact. As more than one business participant in our studies argued,  ``if a business process is critical, we should never accept a technical debt in it, but business pressure is huge''.

Note that currently the framework does not consider the technical debt's type or complexity for the prioritization. A simple ``documentation debt'' will be grouped with a complex ``architectural debt'' if both affect operational IT assets that support core business processes. The stakeholders must decide which item should be prioritized (among those which have higher business impact).

We can also see the potential business impact of technical debt items. For example, if one identifies a technical debt item which affects the sales volume or the system availability, then this item is a candidate to be paid immediately. If one identifies a technical debt item which affects cost (e.g., an inefficient algorithm which increases the cost of cloud infrastructure), the debt could be scheduled to be paid in the short-term. The third impact level (long-term) shows that debt should not ``sleep'' forever (e.g., an architectural issue which may require a considerable time and money investment to be paid).

\section{Framework Evaluation}

The framework was evaluated by applying the following interactions with group B5 (cf.~Table~\ref{table.dsr.phases}):

\begin{enumerate}
  \item A focus group with the whole team (POs, tech leaders, and developers) to discuss the concept of technical debt and its types, to reach a common understanding.
  \item A focus group with technical participants to analyze the current technical debt list. The meeting resulted in a list of technical debt items with affected configuration items. 
  \item A focus group with the technical leaders to identify the configuration items affected by the technical debt items and their dependencies.
  \item A focus group with business professionals to identify the BPs and IT assets which are in their scope.
  \item Two focus groups (two meetings) with business professionals to discuss the business value provided by the BPs and IT assets. The result of this meeting was a list of metrics that must be considered when prioritizing technical debt.
  \item Finally, a meeting with one IT leader and two business participants to analyze the prioritized technical debt list and evaluate the framework.
\end{enumerate}

The final version of the framework was constructed considering the information, discussions, examples, and the experience obtained from interacting with a set of dozens of professionals in 12 different groups from three companies. There were numerous iterations over six months. Each group had its business processes, IT assets, and business metrics. Naturally, many business metrics are shared by many groups (e.g., availability, revenue, transaction volume) while others were specific (e.g., bouncing rate, mean time to rendering). 

During the evaluation, we presented participants with the framework structure (Figure~\ref{fig_tracy}) and reviewed all concepts, extensions, inputs, and outputs with them. Then we presented each artifact discussed during the previous meetings. There were 18 business processes, eight core BPs, and ten other BPs, supported by 5 IT assets (3 operational and 2 to-be operational). Nine business metrics related to the IT assets and business processes, and there were 26 technical debt items of different types (e.g., database, security, UX, architectural, tests, monitoring, code). 

Participants agreed with the final prioritization and gave positive feedback on the metrics related to each item. We asked two questions:

\paragraph{``What did you like and dislike about the framework and the whole experience?''} Participants liked the new perspective on making prioritization decisions. They agreed that with the business information related to technical debt, it could be easier to argue with managers and customers to prioritize critical technical debt. They also liked the idea of having a standard set of business metrics to define ``what is important'' in terms of which technical debt should be prioritized. On the other hand, participants argued that the work to identify business metrics is non-trivial, and they suggested future work to develop a process for the identification of appropriate business metrics. Another challenge that was raised is how to keep metrics and their values up-to-date. This problem also applies to the prioritization canvas (Figure~\ref{fig_prioritization_canvas}), since a ``to-be-operational'' IT asset could move to ``operational'', with implications for the priorities of the corresponding debt items.

\paragraph{``Would you use the framework in the future? Why? Why not?''} All participants agreed with the usefulness of the presented framework and intend to use it in the future to support decision making. A tool is under development to integrate the framework with development tools such as issue trackers and code repositories.

Despite the large number of scenarios from which the framework was conceived, we cannot claim its generalization based on the work done so far. This was the first step of the evaluation process, which must involve more different groups with different business and technical scenarios.

\section{Related work}
Several secondary~\cite{AMPATZOGLOU:2015,Ribeiro:2016,FERNANDEZSANCHEZ201722,terese2019} and tertiary~\cite{RIOS2018-tertiary} studies analyze technical debt research. With regard to technical debt prioritization, it is a common finding that the criteria, tools, and approaches used to prioritize technical debt lack a business perspective. Lenarduzzi et al.~\cite{terese2019} conducted a systematic literature review on technical debt prioritization and identified only three papers that use business-related constraints. They highlight that based on most surveys conducted with practitioners, customer and business factors are the most important to consider when prioritizing technical debt. However, only a few papers addressed such factors.

Ribeiro et al.~\cite{Ribeiro:2016} identified 14 decision-making criteria that can be used by development teams to prioritize the payment of technical debt items but only one of them considers the business aspect of cost-benefit analysis. Ramasubbu and Kemerer \cite{Ramasubbu2019} proposed a three-step normative process framework that incorporates steps for managing technical debt in commercial software development. The process is aligned to PMBOK practices and considers the cost of quality metrics and risk of financial loss as business impact. Different to our approach, they do not use a business process or a wider business-value perspective.
\section{Conclusion and Future Work}

We have presented Tracy, a decision-making framework that prioritizes technical debt considering how IT assets support a company's business processes, thus providing a new perspective on technical debt management. Information about the potential business impact of each technical debt item is crucial to support decisions among stakeholders with different roles. Tracy was constructed using Design Science Research~\cite{dsr.book}, with the participation of 49 practitioners over six months. 

Future work includes building a method to guide the identification of business metrics which support the prioritization. We also plan to create mechanisms to differentiate technical debt items between different business metrics, and we are currently developing a tool to evaluate the framework in more scenarios and companies. The framework also opens up opportunities to advance research on the quantification of interest and principal of technical debt.

\textbf{Acknowledgements.} This research was partially funded by INES 2.0, FACEPE grants PRONEX APQ 0388-1.03/14 and APQ-0399-1.03/17, CAPES grant 88887.136410/2017-00, and CNPq grant 465614/2014-0, and Coordenação de Aperfeiçoamento de Pessoal de Nível Superior - Brasil (CAPES) - Finance Code 001.

\bibliographystyle{IEEEtran}
\bibliography{bibliography/techdebt}

\end{document}